\documentclass[a4paper,fleqn,usenatbib]{mnras}

\usepackage{newtxtext,newtxmath}
\usepackage{mathptmx}
\usepackage{txfonts}

\usepackage[T1]{fontenc}
\usepackage{ae,aecompl}

\usepackage{graphicx}	% Including figure files
\usepackage{amsmath}	% Advanced maths commands
\usepackage{amssymb}	% Extra maths symbols

\title[Are there GC-like abundance patterns in YMCs?]{Searching for GC-like abundance patterns in young massive clusters II. - Results from the Antennae galaxies}
\author[C.~Lardo et al.]{C. Lardo$^{1, 2}$\thanks{E-mail:
C.~Lardo@ljmu.ac.uk}, I.~Cabrera-Ziri $^{1, 3}$, B.~Davies$^{1}$ and N.~Bastian$^{1}$\thanks{Based on observations collected at the European Organisation for Astronomical Research in the Southern Hemisphere under ESO programme 093.B-0023(A).}\\
$^{1}$ Astrophysics Research Institute, Liverpool John Moores University, 146 Brownlow Hill, Liverpool, L3 5RF, UK\\
$^{2}$ Laboratoire d'astrophysique, École Polytechnique Fédérale de Lausanne (EPFL), Observatoire, 1290, Versoix, Switzerland\\
$^{3}$ European Southern Observatory, Karl-Schwarzschild-Stra\ss e 2, D-85748 Garching bei Munchen, Germany}

\date{Accepted XX . Received XX; in original form XX}
\pubyear{2015}

\begin{document}
\label{firstpage}
\pagerange{\pageref{firstpage}--\pageref{lastpage}}
\maketitle

\begin{abstract}
The presence of multiple populations (MPs) with distinctive light element abundances is a widespread phenomenon in clusters older than 6 Gyr. Clusters with masses, luminosities, and sizes comparable to those of ancient globulars are still forming today. Nonetheless, the presence of light element variations has been poorly investigated in such young systems, even if the knowledge of the age at which this phenomenon develops is crucial for theoretical models on MPs. We use $J$-band integrated spectra of three young (7-40 Myr) clusters in NGC~4038 to look for Al variations indicative of MPs. Assuming that the large majority ($\geq$70\%) of stars are characterised by high Al content -- as observed in Galactic clusters with comparable mass; we find that none of the studied clusters show significant Al variations. Small Al spreads have been measured in all the six young clusters observed in the near-infrared. While it is unlikely that young clusters only show low Al whereas old ones display different levels of Al variations; this suggests the possibility that MPs are not present at such young ages at least among the high-mass stellar component.  Alternatively, the fraction of stars with field-like chemistry could be extremely large, mimicking low Al abundances in the integrated spectrum. 
 Finally, since the near-infrared stellar continuum of young clusters is almost entirely due to luminous red supergiants, we can also speculate that MPs only manifest themselves in low mass stars due to some evolutionary mechanism.

%However, such prevalence of stars with pristine composition is not observed in in Galactic old clusters with comparable masses.  

%While studies on young clusters are mandatory to decide between the two envisaged scenarios, theoretical models aimed at explaining the origin of multiple stellar populations in clusters should be able to reproduce the evidence coming from observations of younger cluster systems.

 \end{abstract}


\section{Introduction}\label{introduzione}
Globular clusters (GCs) have been largely demonstrated to host multiple populations (MPs)  with distinctive chemistry \citep[e.g.,][]{grattonREV}. In particular, nearly all GCs show significant spread in their light element content, with strong anti-correlations between C, N, O, Na, Al and He star-to-star differences \citep[e.g.,][]{carretta2009,car09b}. 
Such star-to-star light element variations are not observed only in Galactic GCs, but they have been found also in extragalactic stellar cluster with ages $\geq$ 6 Gyr  \citep[e.g.][]{colucci09,mucciarelli09,dalessandro16,hollyhead16,florian121}.
Conversely, massive ($\sim$10$^{5}$~M$_{\odot}$), intermediate-age clusters (1-3 Gyr) do not show any inhomogeneity in their [Na/Fe] and [Al/Fe] content or photometric splits along their red giant branches (RGBs) when imaged in special filters used to pinpoint the presence of MPs (\citealp{mucciarelli08, mucciarelli14}, Martocchia et al. 2017). 
Finally,  Milky Way (MW) field stars generally do not show anti-correlations (\citealp{martell11}, but see also \citealp{schiavon16}) and no unmistakable chemical anomalies were found in open clusters  \citep[e.g.,][]{bragaglia12}.

The popular scenarios introduced to explain the presence of MPs in GCs generally invoke subsequent episodes of star formation
\citep[e.g.][]{decressin09,dantona16} where Na-poor/O-rich/C-rich stars are the first stars that formed, while Na-rich/O-poor/N-rich stars formed some tens/hundreds of Myr later from the freshly synthesised material from the first generation. Nonetheless, all the proposed models struggle in reproducing the observational evidence \citep[e.g.][]{larsen12,larsen14,kruijssen15,nateHE,bastian15}; suggesting that also other solutions should be investigated to explain this puzzling phenomenon.

The recent discovery that young massive clusters (YMCs) with masses and sizes similar to those of  the ancient GCs  are still forming today offers a fresh perspective to study MPs \citep[e.g.][]{bastian13,bastian14frms,cabreraziri14,cabreraziri15,florianAge}. Such clusters reside in galaxies characterised by strong star formation activity, and  they are so luminous that can be observed also in distant galaxies \citep[e.g.][]{portegies10}. 

Abundance analysis from integrated spectra generally requires population synthesis to define the dominant contribution to the
stellar luminosity. However,  the near-infrared stellar continuum of YMCs is almost entirely due to luminous
red supergiant stars (RSGs) which contribute up to $\sim$95\% of the total flux as soon as YMCs are older than $\sim$7 Myr \citep{gazak14}. 
This allows a major simplification in population synthesis techniques,
making the interpretation of integrated YMC spectra straightforward. Indeed, RSG effective temperatures are constant within $\pm$200 K \citep{davies13} and do not depend on the metallicity  \citep{gazak15}. Also, RSGs in a cluster have similar luminosities and nearly identical masses, and therefore similar surface gravities. Hence, spectra of cluster RSG stars are virtually identical in the $J$-band and their summed spectrum can be modelled with an equivalent, average RSG spectrum. As a matter of fact, such approach has been successfully used to measure metallicities an abundances of YMCs in external galaxies \citep[e.g.][]{larsen06,larsen08,gazak14,lardoANT}.
 
Interestingly, the $J$-band also covers two prominent Al lines. Al is an element that varies in GCs, and 
if GCs and YMCs are indeed the same stellar systems observed at different ages, we would expect to observe such Al variations also in the latter. 
This opens a unique window to quantitatively investigate the presence of light element variations in stellar systems that could become GCs 
in a few Gyr. Following this line of investigation, \citet{cabreraziri16} presented a differential analysis  between Al lines from the integrated $J$-band spectrum of NGC 1705: 1,  a young ($\sim$ 15 Myr) and massive ($\sim$10$^{6}$ M$_{\odot}$) cluster in NGC~1705, and the composite spectrum of field RSGs of similar metallicities, which do not show any light element variation. They exclude at high confidence extreme [Al/Fe] enhancements ($\geq$0.7 dex) like the ones observed in GCs like NGC~2808 and NGC~6752 \citep[e.g.][]{carretta2009}; but they cannot exclude that smaller Al variations are in place.

In \citet{lardoANT} we targeted three YMCs in NGC~4038, a galaxy part of the Antennae merging systems, to 
directly measure their metallicities of  from low-resolution (R [$\lambda/ \Delta \lambda$] $\simeq$ 3200) KMOS spectra. The analysed spectra also include the above-mentioned Al doublet. Here we carry out an analysis similar to that presented in \citet{cabreraziri16} to look for Al variations in these three 
clusters and enlarge the sample of YMCs where Al variations have been constrained. If Al variations are found, we can conclude that YMCs and GCs are the same kind of stellar objected and thus every scenario proposed to explain MPs should be able to fit the observational properties of YMCs
(i.e., no evidence of ongoing star formation or gas reservoirs within YMCs, no evidence of age spreads in YMCs; see \citealp{bastian13,bastian14,bastian14frms,cabreraziri14,cabreraziri15,florianAge}). Conversely, if Al spreads are not present in YMCs we should explore other solutions to explain why MPs are observed only in $\geq$ 6 Gyr old clusters; i.e. in stars with a mass $\leq$ 1.2 M$_{\odot}$.
In both cases, the constrain of the age (at thus the stellar mass) at which MPs are observed turns out to crucial for cluster studies. 
Indeed, the exploration of MPs in YMCs and intermediate age clusters constitutes at the present the only 
way to get fundamental insights in MP origin and not only on MP phenomenology.

This article is structured as follows:
we describe the observational material and data in Section~\ref{OSSERVAZIONI}. We compare observed spectra with synthetic ones 
in Section~\ref{DISCUSSION} in order to look for Al variation. We summarise our findings and 
draw our conclusions in Section~\ref{CONCLUSIONI}.

\begin{table}
\begin{center}
\setlength{\tabcolsep}{0.15cm}

\caption{SSCs targeted. Identification, ages, and masses are from \citet{whitmore10}.
Atmospheric parameters used to measure Al abundances along with [Z/H] values are from \citet{lardoANT}.\label{TAB1}}
\label{SAMPLE}

\begin{tabular} {ccccccc}
\hline
\hline
ID      &   log ($\tau$/yr) & Mass & T$_{\rm{eff}}$   & $\log$ g  & $\xi$  & [Z]  \\
          &                            & (M$_{\sun})$  &(K)      &  (dex)     & (km/s)   & (dex) \\
 \hline
35897 & 7.6	& 4.5 $\times$10$^5$  & 3890  & 	0.4  &1.7  &0.01 $\pm$ 0.07   \\
36731 & 7.6	        & 1.1 $\times$10$^6$  & 3750  & 0.6   &2.0  &0.09 $\pm$ 0.08 \\
50776 &  6.8	& 1.1 $\times$10$^6$   & 3770 & 0.5   &2.0  &0.11 $\pm$ 0.07 \\

\hline
\end{tabular}
\end{center}
\end{table}

\section{Observational material and previous results from literature}\label{OSSERVAZIONI}

Our sample consists of the three YMCs presented in in \citet{lardoANT}.
Their masses, metallicities, and ages are listed in Table~\ref{TAB1} along with other useful information. While we refer to \citet{lardoANT} for a detailed discussion, in the following we briefly summarise data reduction and analysis.

Clusters were observed with VLT/KMOS (PI: Kudritzki: 093.B-0023) 
for a total exposure time of 6000s, yielding to a signal-to-noise ratio $\geq$100 for each spectrum.
KMOS is a spectrograph equipped with 24 integral-field units (IFUs)
that can be allocated within a  7.2\arcmin~diameter field-of-view. (FoV).
Each IFU comprises a projected area on the sky of about 2.8\arcsec $\times$ 2.8\arcsec, which is sampled by an array of 
14$\times$14 spatial pixels with an angular size of 0.2\arcsec~each.
To reduce data, we used the standard recipes provided by the Software Package for Astronomical Reduction with KMOS (SPARK; \citealp{spark}). KMOS IFU data cubes were flat  fielded, wavelength calibrated, and telluric corrected\footnote{We note that the two prominent Al lines investigated in this paper are not affected by strong telluric absorption.} using the
standard KMOS/esorex routines. Additionally, spectra were corrected for variations in spectral resolution and wavelength calibration across 
the field-of-view of each IFU \citep[see][ for details]{lardoANT}. 

\begin{figure}
\includegraphics[width=1.13\columnwidth]{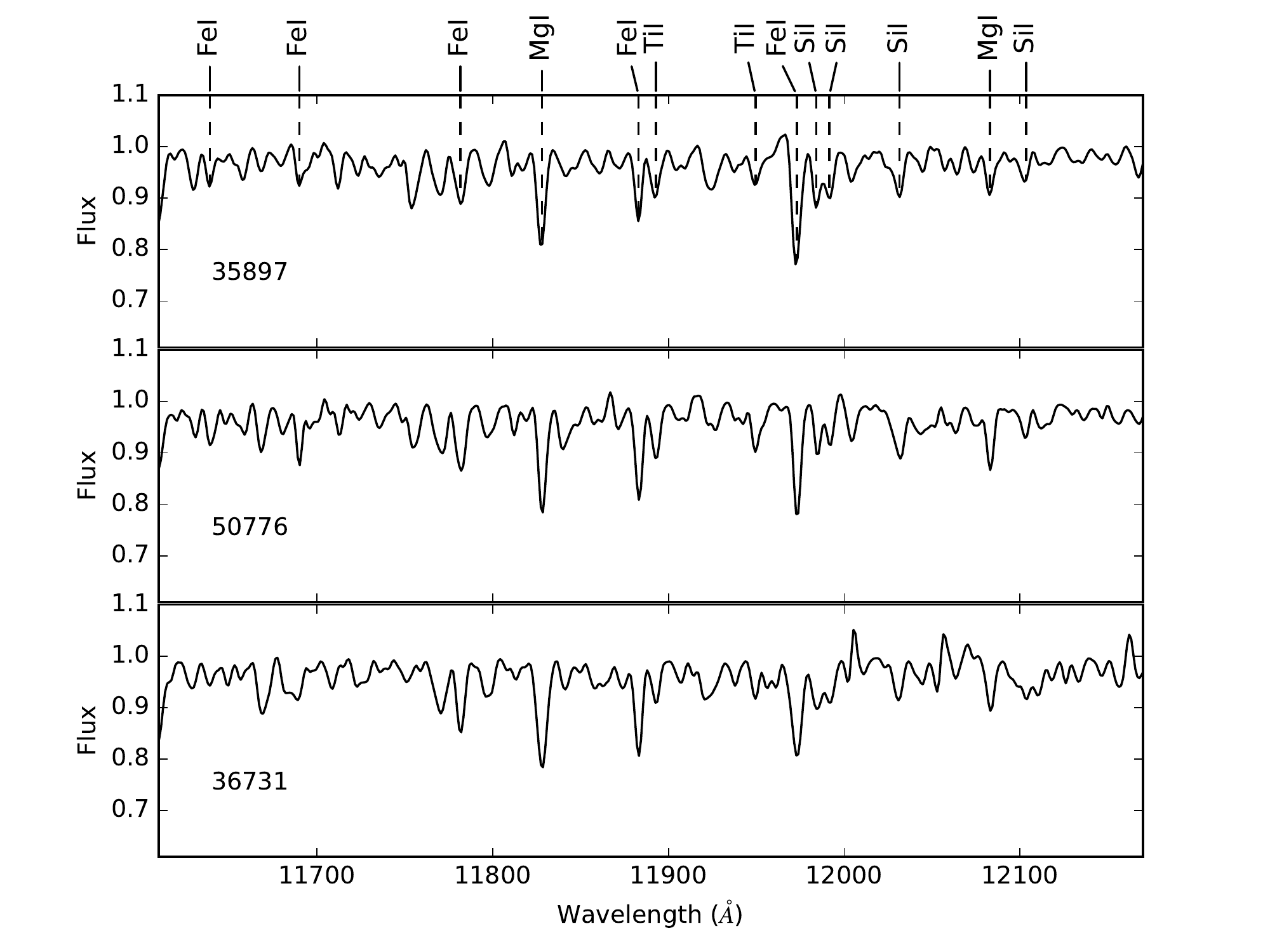}
\caption{Normalised spectra of the analysed clusters in a narrow spectral window in the $J$ band. The metallic absorption lines used  to
measure the overall metallicity [Z] are shown \citep[see][]{lardoANT}. }
        \label{SPECTRA}
   \end{figure}

Metallicity [Z] (normalised to Solar values, [Z] = log Z/Z$_{\sun}$)\footnote{Fe, Mg, Si, and Ti abundances from individual lines 
are assumed to be representative of the metallicity Z.} and atmospheric parameters 
were derived by comparing the observed spectra with a grid of single-star synthetic spectra
degraded at the same spectral resolution than the observed ones.
Model atmospheres were calculated with the MARCS code \citep{gustafsson08}, while the 
synthetic spectra were computed using the updated version of the SIU code \citep{bergemann12}.
Departure from local thermodynamic equilibrium for \ion{Fe}{I},  \ion{Mg}{I}, \ion{Si}{I}, and \ion{Ti}{I} lines were also included \citep[e.g.][]{bergemann12}. The best fit model has been derived through a $\chi ^{2}$ minimisation between the observed spectrum and a
template spectrum at each point in the model grid, taking into accounts possible 
shifts and variations in instrumental spectral resolution between the data and the models.
%We refer again to \citet{lardoANT} for a thoroughly discussion on the adopted method.
In Figure~\ref{SPECTRA} we present the normalised spectra in a narrow region in the $J$-band where most of the
metallic lines are located to illustrate the quality of the data.

\subsection{Aluminium variations in Galactic globulars: data from literature}\label{literature}

Aluminium, along with other light-elements like C, N, O, and Na has been observed to vary in GCs \citep{grattonREV}.
However, while some clusters display large [Al/Fe] variations (which can amount up to $\geq$1 dex),
stars in other clusters share the same Al abundance.

In \citet{cabreraziri16} we draw a homogenous catalogue of literature  [Al/Fe] abundances for 25 GCs in order to quantitatively measure the level of Al enhancement observed in GC stars. For each cluster in our sample we compute $\Delta$[Al/Fe], defined as the difference between the mean [Al/Fe] abundance and the pristine [Al/Fe] abundance in cluster stars; i.e.  $\Delta$[Al/Fe]= mean([Al/Fe])-min([Al/Fe]). We then divide the [Al/Fe] spread observed in GCs in three broad ranges: moderate, intermediate and extreme  according to their $\Delta$[Al/Fe] values ($\Delta$[Al/Fe]= 0.1, 0.3 and 0.7 dex, respectively. We refer to \citet{cabreraziri16} for a complete summary of the measured Al spreads, their standard deviation and maximum [Al/Fe] variation, along with the references to the literature employed to characterise the spectroscopic  sample.

No statistically significant correlation is found between the level of Al enhancement and metallicity (see \citealp{cabreraziri16} and lower panel of Figure~\ref{SPREAD}), suggesting that metal-rich clusters do not necessarily show smaller Al overabundances than metal-poor ones. This 
means that high levels of Al enrichments could potentially also be observed at solar and slightly super-solar metallicities, i.e. the metallicity regime characteristic of YMCs (see Table~\ref{SAMPLE}).

   \begin{figure}
  \centering
\includegraphics[width=0.75\columnwidth]{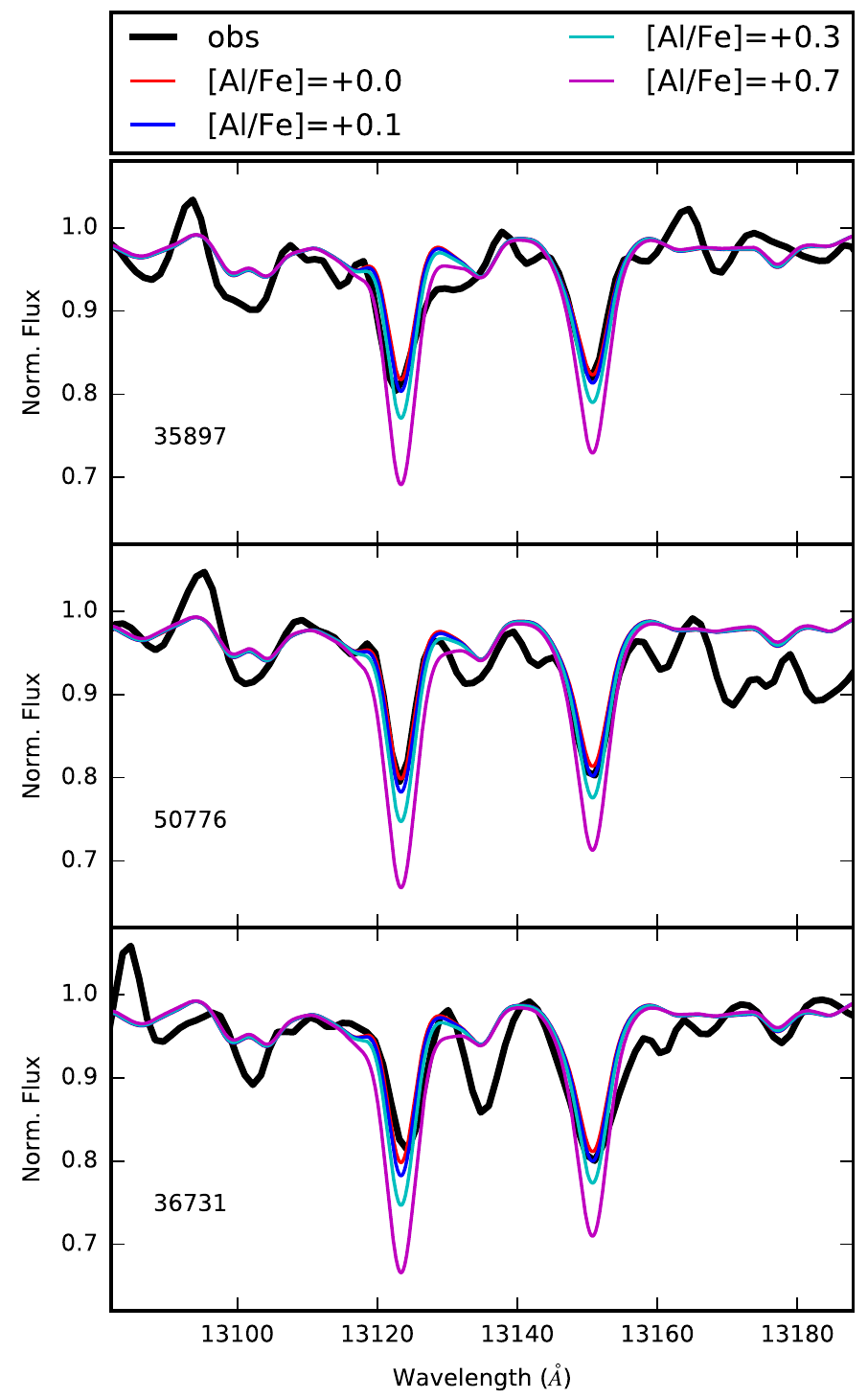}
\caption{Clusters's spectra around the Al doublet at $\sim$ 13123.38 and 13150.71~\AA.
The observed spectra are plotted in black. Superimposed are synthetic spectra with solar-scaled composition
computed with the appropriate atmospheric parameters and metallicities listed in Table~\ref{TAB1}. 
Synthetic spectra are computed by varying the Al abundance by [Al/Fe]= +0.0, +0.1, +0.3, +0.7 dex 
(see legend in the top panel).
}
        \label{AL}
   \end{figure}

\section{Inferring Aluminium abundances from Young Massive cluster integrated spectra}\label{DISCUSSION} 
The KMOS spectra of the three Antennae clusters studied in \citet{lardoANT}. 
include the strong Al doublet at $\simeq$ 13123.38 and 13150.71~\AA.
Hence, the direct comparison of the observed spectra with synthetic ones computed with appropriate atmospheric parameters and metallicities and different Al enhancements can give us indication whether or not MP are present in YMCs \citep[e.g.][]{cabreraziri16}.

If MP light element variations are originated via stellar nucleosynthesis in the interiors of intermediate-to-massive 
stars from a previous generations, larger polluter masses are required to produce Al variations than those needed to originate N and 
Na spreads\footnote{The temperatures required to activate the MgAl cycle are 
T $\geq$70 $\times$ 10$^{6}$ K, while those required to activate the ON and NeNa cycles to produce N and Na variations are T $\geq$40 $\times$ 10$^{6}$ K  \citep[e.g.][]{denisenkov,langer93}.}. This means that YMCs could still display N or Na variations while having 
constant star-to-star Al content. Unfortunately, there are no Na lines in the $J$-band spectra
of RSG, and most of the N and O is locked up in molecules at the temperatures typical of RSG stars and hence is difficult to measure directly. 
While high-resolution optical studies are still possible but extremely time demanding in terms of observing time for such distant systems, they also 
require population synthesis to define the dominant contribution to the stellar luminosity at these wavelengths.
Nonetheless, Al variations are observed in Galactic globulars with masses comparable to those of the YMCs studied here (see Section~\ref{massamet} and \citealp{cabreraziri16}). Thus, there is no reason why a large N and Na enhancement should be associated with no Al variations in YMCs, as Al abundance variations can safely be considered as signature of MPs.

Figure~\ref{AL} shows the normalised integrated light spectra in a region centred around the two strong Al
features. Synthetic spectra with Al abundance varied by +0.0, +0.1, +0.3, and +0.7 dex are also shown; i.e. simulating the enhancement observed in Galactic GC stars with no Al variations or moderate, intermediate, and extreme [Al/Fe] enhancement (see Section~\ref{literature}).
For our computations, we assume that stars with pristine composition in these young and metal-rich systems have solar Al abundance.  
This seems to be supported by observations of field and metal-rich cluster stars \citep[e.g.][]{reddy03,fulbright07,schiavon17} , as well as 
from sparse Al measurements in the Magellanic Clouds \citep[e.g.][]{mucciarelli09,dalessandro16}.
The spectra are all consistent with no Al enhancement, suggesting that chemical anomalies are not present {\em (a)} at such early ages and/or {\em (b)} in massive RSG stars with typical masses $\geq$15 M$_{\odot}$, or {\em (c)} the number of RSGs with pristine chemical composition is significantly larger than that of RSGs showing high Al content. However, significant variations in Al can still be in place if stars with field-like composition (i.e. pristine stars) in the sampled clusters have an depleted Al abundances, i.e. [Al/Fe] $<$ 0.0 dex  because of the chemical evolution of the  Antennae.

   \begin{figure}
    \centering
\includegraphics[width=0.85\columnwidth]{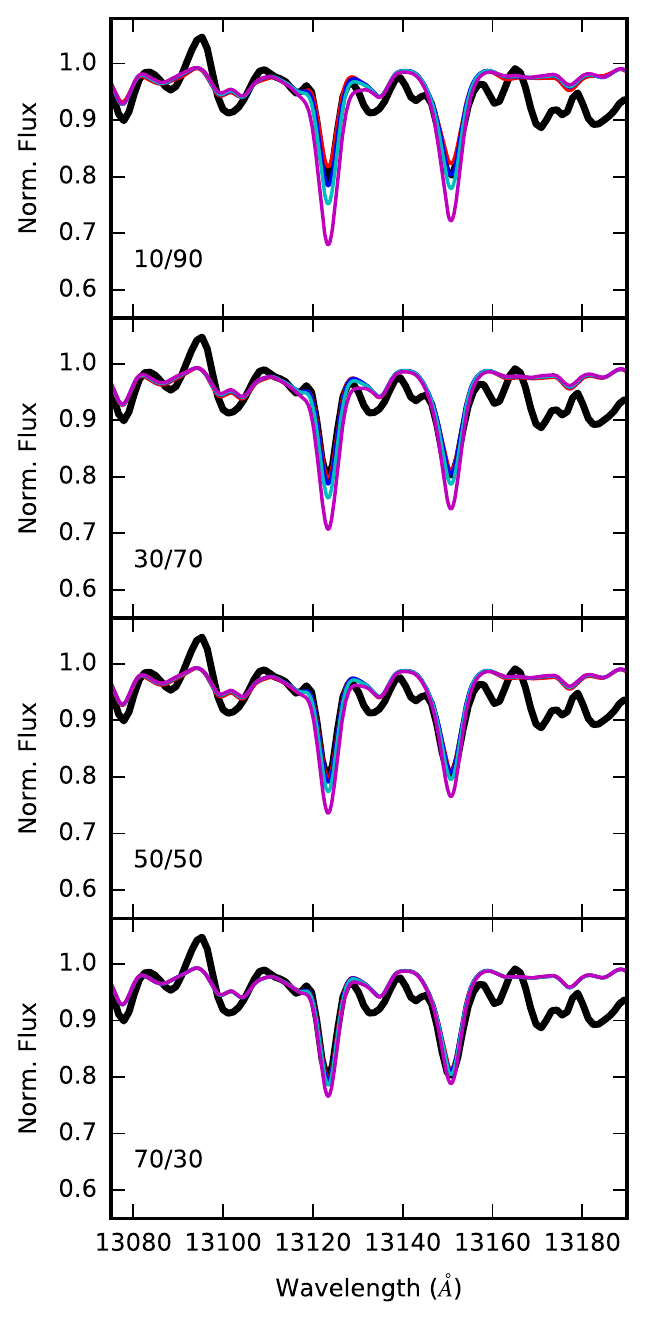}
\caption{KMOS spectra around the $J$-band Al doublet for cluster 50776.
The observed spectra are plotted in black. Superimposed are the composite synthetic spectra 
computed for three different primordial-to-enriched population ratios (10/90, 30/70, 50/50, 70/30; from top to bottom) and Al abundance altered by 
[Al/Fe]= +0.0, +0.1, +0.3, +0.7 dex. Colours are the same as in Figure~\ref{AL}.
}
        \label{EXPER}
  \end{figure}

\subsection{The impact of primordial/enriched ratio on the inferred Al abundances}\label{esperimento}

The integrated spectrum of RSG stars with {\em mixed} chemical composition (primordial + enriched populations, each accounting for a variable and unknown percentage of the whole cluster population) would show weaker Al lines if the number of RSG stars with primordial composition outnumber the number the stars with high Al content. We perform a simple experiment to assess the impact of different primordial-to-enriched population ratios 
on Al doublet appearance in integrated spectra. To this end, we calculate the composite spectrum from a population of 
100 RSG stars with  90\%, 70\%, 50\%, and 30\% of stars showing  moderate, intermediate, and extreme Al enhancement.

Results for clusters 50776 are shown in Figure~\ref{EXPER}. From this Figure it is clear how a small fraction of enriched stars relative to primordial ones could mimic  the absence of any Al enhancement, as the summed spectrum is dominated by RSGs with field-like [Al/Fe] abundances (see bottom panel of Figure~\ref{EXPER}). A quick comparison between observed and theoretical spectra indicates that we would be able to detect only a very large Al enhancement ($\Delta$[Al/Fe]=+0.70 dex) if Al-rich RSGs only account for 30\% of the total RSG population. We would not measure even such large [Al/Fe] variations if enriched RSGs were less then the 30\% of the total RSG population.
On the contrary, we can exclude both intermediate and extreme Al enrichment if SP RSG stars constitute more than 70\% of the RSG population; i.e. $\Delta$[Al/Fe] $\leq$ 0.1 dex.

\begin{figure}
    \centering
\includegraphics[width=\columnwidth]{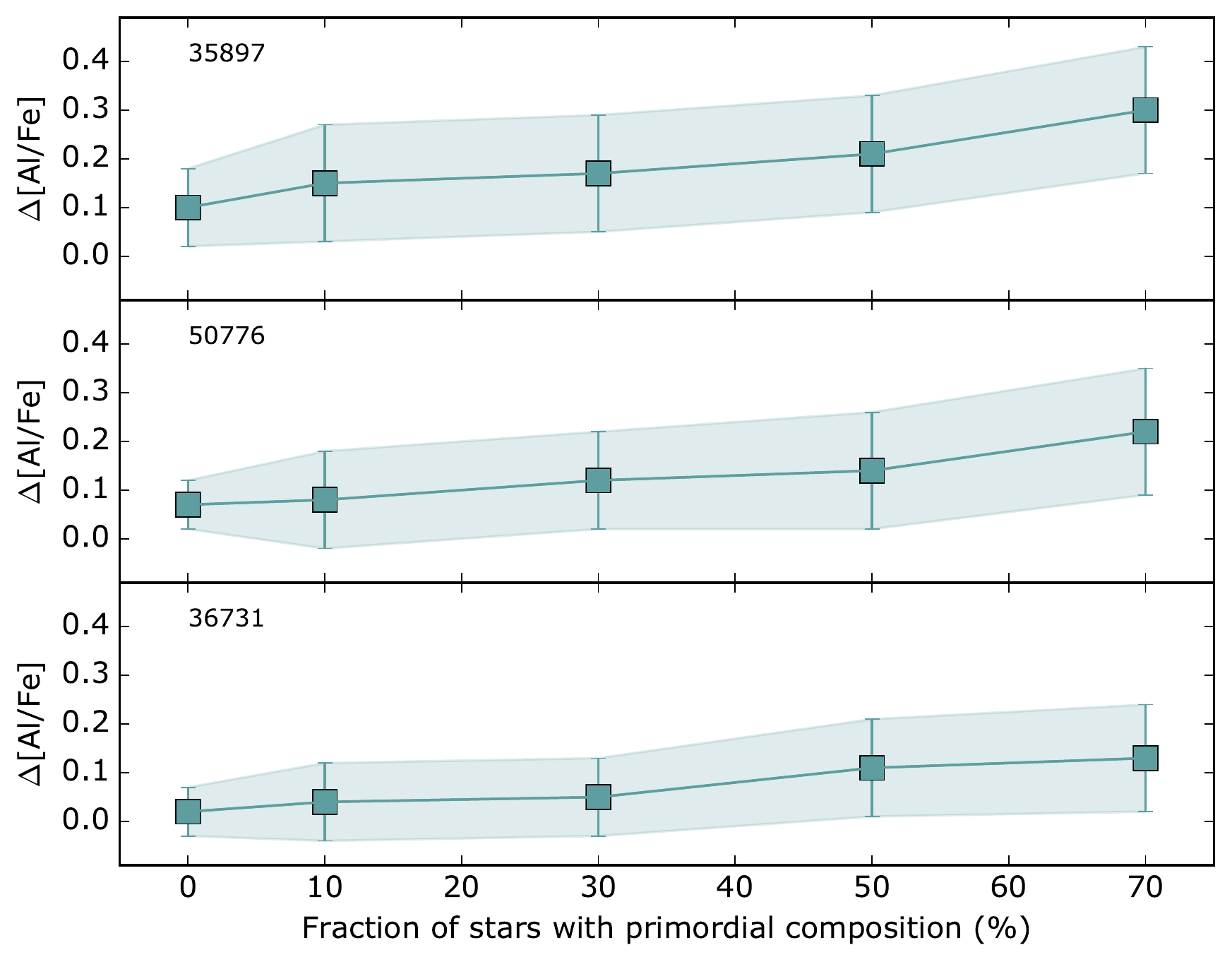}
\caption{ The run of $\Delta$[Al/Fe] spread against the assumed fraction of stars with primordial chemistry is shown for the three analysed clusters. The shaded regions display the 1-$\sigma$ errors on [Al/Fe] measurements.
}
        \label{RATIO}
  \end{figure}

To accurately quantify how different primordial/enriched ratios affect the inferred Al variations, we re-derive abundances (along with their associated 1-$\sigma$ uncertainties) assuming a primordial/enriched ratio equal to 0/100, 10/90, 30/70, 50/50, 70/30.
Figure~\ref{RATIO} shows the run of $\Delta$[Al/Fe] against the assumed fraction of stars with pristine composition. This Figure shows that the inferred [Al/Fe] spread strongly changes when assuming different primordial/enriched mixtures. For example, in the case of 35897, we measure $\Delta$[Al/Fe]=0.30 dex ($\sigma$=0.13 dex) when primordial/enriched=70/30 and a much lower Al spread ($\Delta$[Al/Fe]=0.10 dex, with $\sigma$=0.08 dex) when enriched RSGs account for 90\% of the total RSG population. For 36731, Al variations do not exceed $\Delta$[Al/Fe]=0.10 dex ($\sigma$=0.11 dex) even if stars with primordial Al abundances  amount to 70\% of the total RSG population.

Some comments about the primordial-to-enriched ratio measured in GCs are necessary. 
The {\em global} primordial/enriched ratio is still poorly constrained, as a robust assessment of the relative fraction between primordial and enriched stars needs necessarily to rely on accurate photometry with large radial coverage, from the centre to the cluster outskirts \citep{dalessandro14,massari16}.
Spectroscopic studies find the primordial/enriched ratio to be almost constant and equal to $\simeq$ 70:30 in Galactic GCs \citep{carrettaGLOBAL,bastian15}. Conversely, Hubble Space Telescope photometric investigations reveal that fraction of primordial stars ranges from $\simeq$ 67\% to $\simeq$8\%~and shows a trend with cluster mass \citep{milone16}.  Hence, in the mass range of our clusters, the fraction of stars with field-like composition is expected to be low, i.e. we should be dominated by enriched stars (see also Section~\ref{massamet}).
Spectroscopic studies are mostly based on FLAMES spectra, thus they  sample preferably the outermost regions of clusters due to fiber positioning problems in the inner parts. The cluster crowded cores are accessible through photometric space-based studies, which however are characterised by a much limited FoV ($\sim$3' $\times$ 3')\footnote{For comparison, FLAMES can access targets over a 25' diameter FoV.}. Moreover, the quoted FP/SP ratio
generally refers to the number  of primordial and enriched stars falling within a the instrument FoV, hence this number reflects a {\em local} property of clusters because different regions are sampled at the same projected distance in GCs with disparate structural properties
(this is particularly true for space-based studies as the one presented by \citealp{milone16}). In those respects, a
homogeneous compilation of primordial/enriched ratio values at different half-light (or core) radii for the clusters studied so far with photometry 
would be most welcome.

\subsection{Constraints on Al variations from RSG dominated clusters}\label{massamet}

The near-infrared stellar continuum of YMCs is almost entirely due to luminous RSGs which contribute more than than 95\% of the 
$J$, $H$, and $K$ flux at $\geq$10 Myr \citep[e.g.][]{larsen06,gazak14}. 
Besides the three clusters presented here and NGC~1705:1  \citep{cabreraziri16}, [Al/Fe] measurements exist
for two young massive clusters from near-infrared spectroscopy: NGC~6946-1447, a young ($\sim$10-15 Myr), and massive (1.7 $\times$ 10$^{6}$M$_{\odot}$) cluster in NGC~6949 \citep{larsen06}; and NGC~1569-B, a $\sim$15-25 Myr massive (4.4 $\times$ 10 $^{6}$M$_{\odot}$) cluster in the dwarf irregular NGC~1569 \citep{larsen08}. 
Both clusters have [Al/Fe] abundance ratios consistent with moderate/intermediate spreads ([Al/Fe]=0.25 $\pm$ 0.18 dex and  0.23 $\pm$ 0.11 dex for  NGC~6946-144 and NGC~1569, respectively) as found for the clusters studied here.

\begin{figure}
    \centering
\includegraphics[width=\columnwidth]{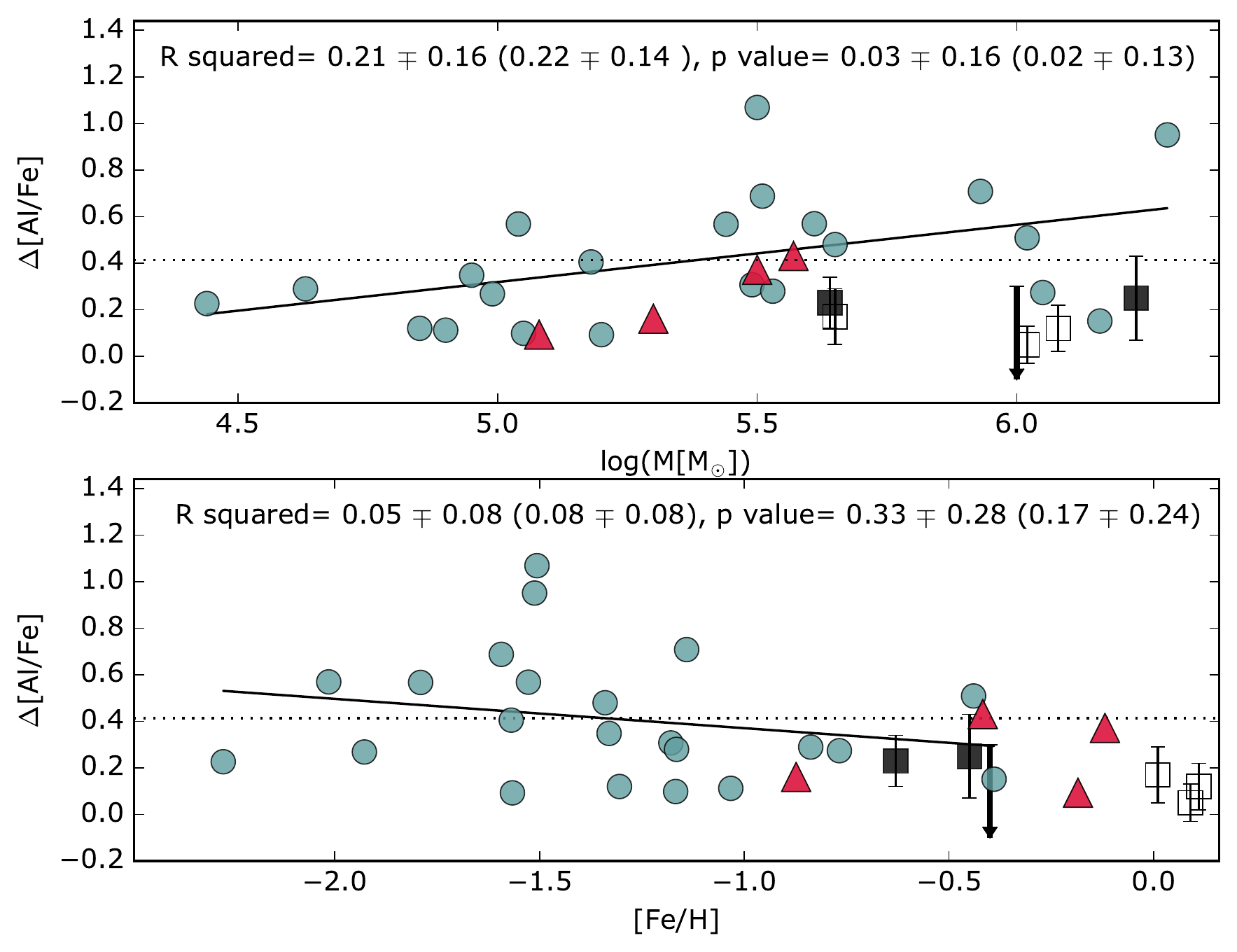}

\caption{ {\em Top panel}: $\Delta$[Al/Fe] as a function of cluster mass. Filled and empty squares (and upper limits) shown the Al spread  derived from integrated light spectra of young clusters \citep{larsen06,larsen08,cabreraziri16} and the three clusters presented in this paper, respectively. {\em Bottom panel:} $\Delta$[Al/Fe] as a function of cluster metallicities.
In both panels teal circles represent the abundances derived by the homogeneous study by Carretta and collaborators. 
Red triangles are metal-rich clusters from the APOGEE survey presented in \citet{schiavon17}.
The solid line is a linear fit to the Carretta data. The dotted line represents the mean $\Delta$[Al/Fe]. The Pearson's correlation coefficient is show on the top-left corners of both panels. The values in parentheses are computed with the inclusion of the metal-rich clusters studied by \citet{schiavon17}. The (anti)correlation between metallicity and $\Delta$[Al/Fe] is mild, while the correlation is weak between $\Delta$[Al/Fe] and [Fe/H]. The p-value is also listed.  Errors on R squared and p values have been computed by means of 5000 bootstrap realisations.
Data are from: \citealp{Carretta:09UVES,Gratton06,Gratton:07GIR,Carretta:14Ter8,Carretta:13N362,Carretta:11N1851,Carretta:14N2808,Carretta:15M80,Carretta:07N6388,Carretta:14N4833,Bragaglia:15,Carretta:10M54,Carretta:12N6752,cabreraziri16,larsen06,larsen08,schiavon17}.
}
        \label{SPREAD}
  \end{figure}

In Figure~\ref{SPREAD}, we plot the observed spread of [Al/Fe] in GCs from the homogeneous survey performed by Carretta and collaborators as a function of the {\em current} cluster mass $M$ with the inclusion of the YMCs studied from integrated light spectroscopy. 
We added also Al measurements obtained by SDSS-III/APOGEE and presented in \citet{schiavon17} for four GCs (namely Palomar~6, Terzan~5, NGC~6553, and NGC~6528) in the Inner Galaxy to probe the high metallicity regime. 
For our YMCs we consider the Al spread derived by assuming that the majority ($\sim$70\%) of RSG stars are enriched in their light element content
(see Figure~\ref{RATIO}), as observed in Galactic old globulars of comparable masses \citep{milone16}.
%However, we stress that for FP/SP ratios equal or smaller than $\leq$ 10/90 we recover much smaller Al variations $\Delta$[Al/Fe] $\leq$ 0.15 dex or less.
Galactic GCs could have lost a large fraction of mass, according to most of the theoretical models invoked to explain multiple 
populations. \citep[e.g.][]{dercole08}. However, all the available observations have called into question the proposed heavy mass-loss \citep[][]{larsen12,bastian15,kruijssen15}. Thus, the assumption that the present-day cluster mass represents a good proxy for the initial one appears reasonable.

The top panel of Figure~\ref{SPREAD} shows a mild correlation between $\Delta$[Al/Fe] and $M$, indicating that the extension (magnitude) of the Al variations is larger for massive clusters \citep[e.g.][]{carretta2009}\footnote{The observed trend possibly reminds 
of the correlation between the masses of GCs and the mean stellar abundances of nitrogen observed by \citet{schiavon13} from integrated spectra of 72 old GCs in the Andromeda galaxy.}.
All the YMCs analysed are located towards large M, yet they all show very small Al enhancement (if any). Hence, even if the Al enhancement measured from integrated light spectra of massive, young, and metal-rich cluster is virtually consistent  with what observed in ancient globulars (see also bottom panel of Figure~\ref{SPREAD}, where we plot the measured $\Delta$[Al/Fe] for Galactic GCs and YMCs studied from integrated light against metallicity), it is not clear why we do not find any YMCs with large Al enhancement since GCs with masses comparable to those of the observed YMCs display both high and low Al enrichment levels. 

This would suggests that MPs are not present among red supergiants in YMCs, which are either too young to display light element variations (assuming that some unknown mechanism is responsible for the onset of MPs only in long-lived objects with masses $\lesssim$1 M$_{\odot}$) or either intrinsically different from GCs where such variation are commonly observed.
Conversely, for some reason variations in Al are indeed present among YMCs stars, but the number of stars showing large Al enhancement only amounts to a small fraction fraction of the total cluster population (see Section~\ref{esperimento}). This could also be the case, even if observational data suggests 
that primordial/enriched ratio anti-correlates with mass. Indeed, according to the relation showed in \citet{milone16}, enriched stars would account for $\sim$80\% of the total number of stars for GCs with masses comparable to those of the three YMCs analysed here.

 \section{Summary and Conclusions}\label{CONCLUSIONI}
 We perform an analysis of integrated $J$-band spectra for three young ($\sim$ 7 -- 40 Myr) and massive (M $\sim$ 5 $\times$ 10$^{5}$ -- 10$^{6}$ M$_{\odot}$) clusters in NGC~4038, part of the Antennae interacting system, in order to look for Al variations. If present, such variations would demonstrate that MPs are present {\em (a)} also in massive RSGs (with typical masses $\geq$ 15 M$_{\odot}$) and/or {\em (b)} since the very early stages of cluster evolution.
 We find that our integrated spectra are consistent with showing only an intermediate [Al/Fe] spread (i.e. $\Delta$[Al/Fe] $\leq$ 0.3 dex) if enriched RSGs represent only a minor fraction ($\leq$ 30\%) of the total RSG total population. If we assume that the vast majority of stars in clusters are enriched stars (as observed in old globulars with comparable mass; e.g. , \citealp{milone16}), we conclude that all clusters are consistent with virtually showing no Al variations, i.e. $\Delta$[Al/Fe] $\lesssim$ 0.15 dex. It should be emphasised that these conclusion are valid only if stars with field stars in the Antennae  have solar [Al/Fe] abundances, so that the possible primordial population has  [Al/Fe] $\simeq$ 0.

 No high resolution Al abundances are available for GCs in the metallicity regime covered by our Antennae clusters ([Fe/H] $\simeq$ 0.0-0.1 dex),  to perform a direct comparison. However, $\sim$ 40\%~of the Galactic GCs show moderate or higher  [Al/Fe] enhancement (see Figure~\ref{SPREAD}~and~\citealp{cabreraziri16}). The six YMCs for which Al abundances exist (i.e. the three presented here and the clusters from \citealp{cabreraziri16} and \citealp{larsen06,larsen08}) are all are consistent with having Al variations $\leq$ 0.3 dex. So the probability that we have missed clusters with Al variation larger that 0.3 dex is P$\sim$0.60$^{6}$ $\leq$ 1\%. {\em This probability is low enough to confidently state that RSG stars in YMCs do not  possess any significant Al overabundance.}
 
 If multiple stellar populations are indeed present in YMCs but their RSGs only show small Al star-to-star variations, it remains to be explained why GCs show different levels of [Al/Fe] spreads and YMCs do not.
On the other hand, the absence of high Al enhancement could also indicate that YMCs do not host MPs. This would imply 
either that  GCs and YMCs are inherently different or MPs only develop later during cluster evolution due to some unknown evolutionary mechanism which affect only low-mass stars with masses $\leq$1 M$_{\odot}$ (i.e. the highest stellar mass at which MP have been detected so far;  see~\citealp{hollyhead16}) and not the more massive RSG stars  
sampled in near-infrared spectroscopic studies of YMCs.

Under the assumption that YMCs represent  the young counterparts of the ancient GCs, we can briefly discuss the implications of our results 
for the most popular multiple population formation scenarios.
The timescales associated with the fast rotating massive stars and asymptotic giant branch (AGB) stars scenarios are $\leq$10 Myr and between 30-200 Myr, respectively. Thus, according to these scenarios, we should be able to observe Al variations in our clusters, which have ages between $\sim$7 - 40 Myr.  However, models where enriched stars forms out of a mixture of pristine material and ejecta from AGB stars with pristine composition
  require that a burst of star formation within clusters form enriched stars with an initial mass function (IMF) truncated at high masses to reduce the required mass budget and avoid more supernovae \citep[e.g.,][]{dercole08,dercole10}\footnote{Stars more massive than 8-9 M$_{\odot}$ would explode as Type II supernovae (SNe II), whereas very small Fe spreads are observed in  GC stars \citep[e.g.][]{carretta2009}. Also, as the first SNe II starts to explode after $\sim$5 Myr, they would sweep out the gas returned by AGB stars -- which have timescales on the order of tens of Myr, preventing the formation of a substantial amount of enriched stars.}.
Therefore, even though it is not clear why enriched stars should be characterised an IMF radically different from that of primordial ones, our result is still compatible with AGB model which assume that no Al-rich  with masses above $\sim$8 M$_{\odot}$ can form within them; i.e. clusters do not form enriched red supergiants.

Finally, while large and homogeneous photometric and spectroscopic surveys \citep[][]{piotto15,car09b,carretta2009} had the merit of having built a benchmark of MP properties in old ($\geq$10 Gyr) clusters with {\em standard} structural and orbital properties, they partially overlooked systems that occupy interesting locations in the parameter space (either in mass, age, metallicity, environment at formation). In these regards, 
fresh insights can be obtained by the study of YMCs or clusters with intermediate-age. While the MW  lacks a population of massive stellar clusters with ages below $\leq$9-10 Gyr, the Magellanic Clouds host significant population of such clusters. A systematic photometric survey of clusters in a large age range  (from 100 Myr to 12 Gyr) using a special combination of filters \citep[e.g.,][]{marino08,lardo11,sbordone11,monelli13,piotto15} can reveal the presence of splits or spreads along the photometric evolutionary sequences indicative of MPs and conclusively demonstrate weather threshold in age or stellar mass exist to observe MPs (e.g.; \citealp{Florian:17b,florian121,hollyhead16}; Martocchia et al. 2017; Hollyhead et al. 2017 submitted).

\section{acknowledgements}
We thank the anonymous referee for a detailed report that
helped to improve the presentation of our results. CL thanks the Swiss National Science Foundation for supporting this research through the Ambizione grant number PZ00P2\_168065. NB gratefully acknowledges financial support from the Royal Society (University Research Fellowship) and the European Research Council (ERC-CoG-646928, Multi-Pop).

\bibliographystyle{mn2e}
\bibliography{bibliography}

\begin{thebibliography}{69}
\expandafter\ifx\csname natexlab\endcsname\relax\def\natexlab#1{#1}\fi

\bibitem[{{Bastian} {et~al}\mbox{.}(2013){Bastian}, {Cabrera-Ziri}, {Davies},
  \& {Larsen}}]{bastian13}
{Bastian} N., {Cabrera-Ziri} I., {Davies} B., {Larsen} S.~S., 2013, MNRAS, 436,
  2852

\bibitem[{{Bastian}, {Cabrera-Ziri} \& {Salaris}(2015){Bastian},
  {Cabrera-Ziri}, \& {Salaris}}]{nateHE}
{Bastian} N., {Cabrera-Ziri} I., {Salaris} M., 2015, MNRAS, 449, 3333

\bibitem[{{Bastian}, {Hollyhead} \& {Cabrera-Ziri}(2014){Bastian}, {Hollyhead},
  \& {Cabrera-Ziri}}]{bastian14frms}
{Bastian} N., {Hollyhead} K., {Cabrera-Ziri} I., 2014, MNRAS, 445, 378

\bibitem[{{Bastian} \& {Lardo}(2015)}]{bastian15}
{Bastian} N., {Lardo} C., 2015, MNRAS, 453, 357

\bibitem[{{Bastian} \& {Strader}(2014)}]{bastian14}
{Bastian} N., {Strader} J., 2014, MNRAS, 443, 3594

\bibitem[{{Bergemann} {et~al}\mbox{.}(2012){Bergemann}, {Lind}, {Collet},
  {Magic}, \& {Asplund}}]{bergemann12}
{Bergemann} M., {Lind} K., {Collet} R., {Magic} Z., {Asplund} M., 2012, MNRAS,
  427, 27

\bibitem[{{Bragaglia} {et~al}\mbox{.}(2015){Bragaglia}, {Carretta}, {Sollima},
  {Donati}, {D'Orazi}, {Gratton}, {Lucatello}, \& {Sneden}}]{Bragaglia:15}
{Bragaglia} A., {Carretta} E., {Sollima} A., {Donati} P., {D'Orazi} V.,
  {Gratton} R.~G., {Lucatello} S., {Sneden} C., 2015, A\&A, 583, A69

\bibitem[{{Bragaglia} {et~al}\mbox{.}(2012){Bragaglia}, {Gratton}, {Carretta},
  {D'Orazi}, {Sneden}, \& {Lucatello}}]{bragaglia12}
{Bragaglia} A., {Gratton} R.~G., {Carretta} E., {D'Orazi} V., {Sneden} C.,
  {Lucatello} S., 2012, A\&A, 548, A122

\bibitem[{{Cabrera-Ziri} {et~al}\mbox{.}(2014){Cabrera-Ziri}, {Bastian},
  {Davies}, {Magris}, {Bruzual}, \& {Schweizer}}]{cabreraziri14}
{Cabrera-Ziri} I., {Bastian} N., {Davies} B., {Magris} G., {Bruzual} G.,
  {Schweizer} F., 2014, MNRAS, 441, 2754

\bibitem[{{Cabrera-Ziri} {et~al}\mbox{.}(2015){Cabrera-Ziri}, {Bastian},
  {Longmore}, {Brogan}, {Hollyhead}, {Larsen}, {Whitmore}, {Johnson},
  {Chandar}, {Henshaw}, {Davies}, \& {Hibbard}}]{cabreraziri15}
{Cabrera-Ziri} I. {et~al.}, 2015, MNRAS, 448, 2224

\bibitem[{{Cabrera-Ziri} {et~al}\mbox{.}(2016){Cabrera-Ziri}, {Lardo},
  {Davies}, {Bastian}, {Beccari}, {Larsen}, \& {Hernandez}}]{cabreraziri16}
{Cabrera-Ziri} I., {Lardo} C., {Davies} B., {Bastian} N., {Beccari} G.,
  {Larsen} S.~S., {Hernandez} S., 2016, MNRAS, 460, 1869

\bibitem[{{Carretta}(2014)}]{Carretta:14N2808}
{Carretta} E., 2014, ApJL, 795, L28

\bibitem[{{Carretta} {et~al}\mbox{.}(2009{\natexlab{a}}){Carretta},
  {Bragaglia}, {Gratton}, \& {Lucatello}}]{carretta2009}
{Carretta} E., {Bragaglia} A., {Gratton} R., {Lucatello} S.,
  2009{\natexlab{a}}, A\&A, 505, 139

\bibitem[{{Carretta} {et~al}\mbox{.}(2009{\natexlab{b}}){Carretta},
  {Bragaglia}, {Gratton}, \& {Lucatello}}]{Carretta:09UVES}
{Carretta} E., {Bragaglia} A., {Gratton} R., {Lucatello} S.,
  2009{\natexlab{b}}, A\&A, 505, 139

\bibitem[{{Carretta} {et~al}\mbox{.}(2014{\natexlab{a}}){Carretta},
  {Bragaglia}, {Gratton}, {D'Orazi}, {Lucatello}, {Momany}, {Sollima},
  {Bellazzini}, {Catanzaro}, \& {Leone}}]{Carretta:14N4833}
{Carretta} E. {et~al.}, 2014{\natexlab{a}}, A\&A, 564, A60

\bibitem[{{Carretta} {et~al}\mbox{.}(2014{\natexlab{b}}){Carretta},
  {Bragaglia}, {Gratton}, {D'Orazi}, {Lucatello}, \&
  {Sollima}}]{Carretta:14Ter8}
{Carretta} E., {Bragaglia} A., {Gratton} R.~G., {D'Orazi} V., {Lucatello} S.,
  {Sollima} A., 2014{\natexlab{b}}, A\&A, 561, A87

\bibitem[{{Carretta} {et~al}\mbox{.}(2015){Carretta}, {Bragaglia}, {Gratton},
  {D'Orazi}, {Lucatello}, {Sollima}, {Momany}, {Catanzaro}, \&
  {Leone}}]{Carretta:15M80}
{Carretta} E. {et~al.}, 2015, A\&A, 578, A116

\bibitem[{{Carretta} {et~al}\mbox{.}(2010{\natexlab{a}}){Carretta},
  {Bragaglia}, {Gratton}, {Lucatello}, {Bellazzini}, {Catanzaro}, {Leone},
  {Momany}, {Piotto}, \& {D'Orazi}}]{Carretta:10M54}
{Carretta} E. {et~al.}, 2010{\natexlab{a}}, ApJL, 714, L7

\bibitem[{{Carretta} {et~al}\mbox{.}(2009{\natexlab{c}}){Carretta},
  {Bragaglia}, {Gratton}, {Lucatello}, {Catanzaro}, {Leone}, {Bellazzini},
  {Claudi}, {D'Orazi}, {Momany}, {Ortolani}, {Pancino}, {Piotto},
  {Recio-Blanco}, \& {Sabbi}}]{car09b}
{Carretta} E. {et~al.}, 2009{\natexlab{c}}, A\&A, 505, 117

\bibitem[{{Carretta} {et~al}\mbox{.}(2012){Carretta}, {Bragaglia}, {Gratton},
  {Lucatello}, \& {D'Orazi}}]{Carretta:12N6752}
{Carretta} E., {Bragaglia} A., {Gratton} R.~G., {Lucatello} S., {D'Orazi} V.,
  2012, ApJL, 750, L14

\bibitem[{{Carretta} {et~al}\mbox{.}(2013){Carretta}, {Bragaglia}, {Gratton},
  {Lucatello}, {D'Orazi}, {Bellazzini}, {Catanzaro}, {Leone}, {Momany}, \&
  {Sollima}}]{Carretta:13N362}
{Carretta} E. {et~al.}, 2013, A\&A, 557, A138

\bibitem[{{Carretta} {et~al}\mbox{.}(2007){Carretta}, {Bragaglia}, {Gratton},
  {Momany}, {Recio-Blanco}, {Cassisi}, {Fran{\c c}ois}, {James}, {Lucatello},
  \& {Moehler}}]{Carretta:07N6388}
{Carretta} E. {et~al.}, 2007, A\&A, 464, 967

\bibitem[{{Carretta} {et~al}\mbox{.}(2010{\natexlab{b}}){Carretta},
  {Bragaglia}, {Gratton}, {Recio-Blanco}, {Lucatello}, {D'Orazi}, \&
  {Cassisi}}]{carrettaGLOBAL}
{Carretta} E., {Bragaglia} A., {Gratton} R.~G., {Recio-Blanco} A., {Lucatello}
  S., {D'Orazi} V., {Cassisi} S., 2010{\natexlab{b}}, A\&A, 516, A55+

\bibitem[{{Carretta} {et~al}\mbox{.}(2011){Carretta}, {Lucatello}, {Gratton},
  {Bragaglia}, \& {D'Orazi}}]{Carretta:11N1851}
{Carretta} E., {Lucatello} S., {Gratton} R.~G., {Bragaglia} A., {D'Orazi} V.,
  2011, A\&A, 533, A69

\bibitem[{{Colucci} {et~al}\mbox{.}(2009){Colucci}, {Bernstein}, {Cameron},
  {McWilliam}, \& {Cohen}}]{colucci09}
{Colucci} J.~E., {Bernstein} R.~A., {Cameron} S., {McWilliam} A., {Cohen}
  J.~G., 2009, ApJ, 704, 385

\bibitem[{{Dalessandro} {et~al}\mbox{.}(2016){Dalessandro}, {Lapenna},
  {Mucciarelli}, {Origlia}, {Ferraro}, \& {Lanzoni}}]{dalessandro16}
{Dalessandro} E., {Lapenna} E., {Mucciarelli} A., {Origlia} L., {Ferraro}
  F.~R., {Lanzoni} B., 2016, ApJ, 829, 77

\bibitem[{{Dalessandro} {et~al}\mbox{.}(2014){Dalessandro}, {Massari},
  {Bellazzini}, {Miocchi}, {Mucciarelli}, {Salaris}, {Cassisi}, {Ferraro}, \&
  {Lanzoni}}]{dalessandro14}
{Dalessandro} E. {et~al.}, 2014, ApJL, 791, L4

\bibitem[{{D'Antona} {et~al}\mbox{.}(2016){D'Antona}, {Vesperini}, {D'Ercole},
  {Ventura}, {Milone}, {Marino}, \& {Tailo}}]{dantona16}
{D'Antona} F., {Vesperini} E., {D'Ercole} A., {Ventura} P., {Milone} A.~P.,
  {Marino} A.~F., {Tailo} M., 2016, MNRAS, 458, 2122

\bibitem[{{Davies} {et~al}\mbox{.}(2013{\natexlab{a}}){Davies}, {Kudritzki},
  {Plez}, {Trager}, {Lan{\c c}on}, {Gazak}, {Bergemann}, {Evans}, \&
  {Chiavassa}}]{davies13}
{Davies} B. {et~al.}, 2013{\natexlab{a}}, ApJ, 767, 3

\bibitem[{{Davies} {et~al}\mbox{.}(2013{\natexlab{b}}){Davies}, {Agudo Berbel},
  {Wiezorrek}, {Cirasuolo}, {F{\"o}rster Schreiber}, {Jung}, {Muschielok},
  {Ott}, {Ramsay}, {Schlichter}, {Sharples}, \& {Wegner}}]{spark}
{Davies} R.~I. {et~al.}, 2013{\natexlab{b}}, A\&A, 558, A56

\bibitem[{{Decressin} {et~al}\mbox{.}(2009){Decressin}, {Charbonnel}, {Siess},
  {Palacios}, {Meynet}, \& {Georgy}}]{decressin09}
{Decressin} T., {Charbonnel} C., {Siess} L., {Palacios} A., {Meynet} G.,
  {Georgy} C., 2009, A\&A, 505, 727

\bibitem[{{Denisenkov} \& {Denisenkova}(1989)}]{denisenkov}
{Denisenkov} P.~A., {Denisenkova} S.~N., 1989, Astronomicheskij Tsirkulyar,
  1538, 11

\bibitem[{{D'Ercole} {et~al}\mbox{.}(2010){D'Ercole}, {D'Antona}, {Ventura},
  {Vesperini}, \& {McMillan}}]{dercole10}
{D'Ercole} A., {D'Antona} F., {Ventura} P., {Vesperini} E., {McMillan}
  S.~L.~W., 2010, MNRAS, 407, 854

\bibitem[{{D'Ercole} {et~al}\mbox{.}(2008){D'Ercole}, {Vesperini}, {D'Antona},
  {McMillan}, \& {Recchi}}]{dercole08}
{D'Ercole} A., {Vesperini} E., {D'Antona} F., {McMillan} S.~L.~W., {Recchi} S.,
  2008, MNRAS, 391, 825

\bibitem[{{Fulbright}, {McWilliam} \& {Rich}(2007){Fulbright}, {McWilliam}, \&
  {Rich}}]{fulbright07}
{Fulbright} J.~P., {McWilliam} A., {Rich} R.~M., 2007, ApJ, 661, 1152

\bibitem[{{Gazak} {et~al}\mbox{.}(2014){Gazak}, {Davies}, {Bastian},
  {Kudritzki}, {Bergemann}, {Plez}, {Evans}, {Patrick}, {Bresolin}, \&
  {Schinnerer}}]{gazak14}
{Gazak} J.~Z. {et~al.}, 2014, ApJ, 787, 142

\bibitem[{{Gazak} {et~al}\mbox{.}(2015){Gazak}, {Kudritzki}, {Evans},
  {Patrick}, {Davies}, {Bergemann}, {Plez}, {Bresolin}, {Bender}, {Wegner},
  {Bonanos}, \& {Williams}}]{gazak15}
{Gazak} J.~Z. {et~al.}, 2015, ApJ, 805, 182

\bibitem[{{Gratton}, {Carretta} \& {Bragaglia}(2012){Gratton}, {Carretta}, \&
  {Bragaglia}}]{grattonREV}
{Gratton} R.~G., {Carretta} E., {Bragaglia} A., 2012, A\&ARv, 20, 50

\bibitem[{{Gratton} {et~al}\mbox{.}(2007){Gratton}, {Lucatello}, {Bragaglia},
  {Carretta}, {Cassisi}, {Momany}, {Pancino}, {Valenti}, {Caloi}, {Claudi},
  {D'Antona}, {Desidera}, {Fran{\c c}ois}, {James}, {Moehler}, {Ortolani},
  {Pasquini}, {Piotto}, \& {Recio-Blanco}}]{Gratton:07GIR}
{Gratton} R.~G. {et~al.}, 2007, A\&A, 464, 953

\bibitem[{{Gratton} {et~al}\mbox{.}(2006){Gratton}, {Lucatello}, {Bragaglia},
  {Carretta}, {Momany}, {Pancino}, \& {Valenti}}]{Gratton06}
{Gratton} R.~G., {Lucatello} S., {Bragaglia} A., {Carretta} E., {Momany} Y.,
  {Pancino} E., {Valenti} E., 2006, A\&A, 455, 271

\bibitem[{{Gustafsson} {et~al}\mbox{.}(2008){Gustafsson}, {Edvardsson},
  {Eriksson}, {J{\o}rgensen}, {Nordlund}, \& {Plez}}]{gustafsson08}
{Gustafsson} B., {Edvardsson} B., {Eriksson} K., {J{\o}rgensen} U.~G.,
  {Nordlund} {\AA}., {Plez} B., 2008, A\&A, 486, 951

\bibitem[{{Hollyhead} {et~al}\mbox{.}(2017){Hollyhead}, {Kacharov}, {Lardo},
  {Bastian}, {Hilker}, {Rejkuba}, {Koch}, {Grebel}, \&
  {Georgiev}}]{hollyhead16}
{Hollyhead} K. {et~al.}, 2017, MNRAS, 465, L39

\bibitem[{{Kruijssen}(2015)}]{kruijssen15}
{Kruijssen} J.~M.~D., 2015, MNRAS, 454, 1658

\bibitem[{{Langer}, {Hoffman} \& {Sneden}(1993){Langer}, {Hoffman}, \&
  {Sneden}}]{langer93}
{Langer} G.~E., {Hoffman} R., {Sneden} C., 1993, PASP, 105, 301

\bibitem[{{Lardo} {et~al}\mbox{.}(2011){Lardo}, {Bellazzini}, {Pancino},
  {Carretta}, {Bragaglia}, \& {Dalessandro}}]{lardo11}
{Lardo} C., {Bellazzini} M., {Pancino} E., {Carretta} E., {Bragaglia} A.,
  {Dalessandro} E., 2011, A\&A, 525, A114+

\bibitem[{{Lardo} {et~al}\mbox{.}(2015){Lardo}, {Davies}, {Kudritzki}, {Gazak},
  {Evans}, {Patrick}, {Bergemann}, \& {Plez}}]{lardoANT}
{Lardo} C., {Davies} B., {Kudritzki} R.-P., {Gazak} J.~Z., {Evans} C.~J.,
  {Patrick} L.~R., {Bergemann} M., {Plez} B., 2015, ApJ, 812, 160

\bibitem[{{Larsen} {et~al}\mbox{.}(2014){Larsen}, {Brodie}, {Forbes}, \&
  {Strader}}]{larsen14}
{Larsen} S.~S., {Brodie} J.~P., {Forbes} D.~A., {Strader} J., 2014, A\&A, 565,
  A98

\bibitem[{{Larsen} {et~al}\mbox{.}(2008){Larsen}, {Origlia}, {Brodie}, \&
  {Gallagher}}]{larsen08}
{Larsen} S.~S., {Origlia} L., {Brodie} J., {Gallagher} J.~S., 2008, MNRAS, 383,
  263

\bibitem[{{Larsen} {et~al}\mbox{.}(2006){Larsen}, {Origlia}, {Brodie}, \&
  {Gallagher}}]{larsen06}
{Larsen} S.~S., {Origlia} L., {Brodie} J.~P., {Gallagher} J.~S., 2006, MNRAS,
  368, L10

\bibitem[{{Larsen}, {Strader} \& {Brodie}(2012){Larsen}, {Strader}, \&
  {Brodie}}]{larsen12}
{Larsen} S.~S., {Strader} J., {Brodie} J.~P., 2012, A\&A, 544, L14

\bibitem[{{Marino} {et~al}\mbox{.}(2008){Marino}, {Villanova}, {Piotto},
  {Milone}, {Momany}, {Bedin}, \& {Medling}}]{marino08}
{Marino} A.~F., {Villanova} S., {Piotto} G., {Milone} A.~P., {Momany} Y.,
  {Bedin} L.~R., {Medling} A.~M., 2008, A\&A, 490, 625

\bibitem[{{Martell} {et~al}\mbox{.}(2011){Martell}, {Smolinski}, {Beers}, \&
  {Grebel}}]{martell11}
{Martell} S.~L., {Smolinski} J.~P., {Beers} T.~C., {Grebel} E.~K., 2011, \aap,
  534, A136

\bibitem[{{Massari} {et~al}\mbox{.}(2016){Massari}, {Lapenna}, {Bragaglia},
  {Dalessandro}, {Contreras Ramos}, \& {Amigo}}]{massari16}
{Massari} D., {Lapenna} E., {Bragaglia} A., {Dalessandro} E., {Contreras Ramos}
  R., {Amigo} P., 2016, MNRAS, 458, 4162

\bibitem[{{Milone} {et~al}\mbox{.}(2017){Milone}, {Piotto}, {Renzini},
  {Marino}, {Bedin}, {Vesperini}, {D'Antona}, {Nardiello}, {Anderson}, {King},
  {Yong}, {Bellini}, {Aparicio}, {Barbuy}, {Brown}, {Cassisi}, {Ortolani},
  {Salaris}, {Sarajedini}, \& {van der Marel}}]{milone16}
{Milone} A.~P. {et~al.}, 2017, MNRAS, 464, 3636

\bibitem[{{Monelli} {et~al}\mbox{.}(2013){Monelli}, {Milone}, {Stetson},
  {Marino}, {Cassisi}, {del Pino Molina}, {Salaris}, {Aparicio}, {Asplund},
  {Grundahl}, {Piotto}, {Weiss}, {Carrera}, {Cebri{\'a}n}, {Murabito},
  {Pietrinferni}, \& {Sbordone}}]{monelli13}
{Monelli} M. {et~al.}, 2013, MNRAS, 431, 2126

\bibitem[{{Mucciarelli} {et~al}\mbox{.}(2008){Mucciarelli}, {Carretta},
  {Origlia}, \& {Ferraro}}]{mucciarelli08}
{Mucciarelli} A., {Carretta} E., {Origlia} L., {Ferraro} F.~R., 2008, AJ, 136,
  375

\bibitem[{{Mucciarelli} {et~al}\mbox{.}(2014){Mucciarelli}, {Dalessandro},
  {Ferraro}, {Origlia}, \& {Lanzoni}}]{mucciarelli14}
{Mucciarelli} A., {Dalessandro} E., {Ferraro} F.~R., {Origlia} L., {Lanzoni}
  B., 2014, ApJL, 793, L6

\bibitem[{{Mucciarelli} {et~al}\mbox{.}(2009){Mucciarelli}, {Origlia},
  {Ferraro}, \& {Pancino}}]{mucciarelli09}
{Mucciarelli} A., {Origlia} L., {Ferraro} F.~R., {Pancino} E., 2009, ApJL, 695,
  L134

\bibitem[{{Niederhofer} {et~al}\mbox{.}(2017{\natexlab{a}}){Niederhofer},
  {Bastian}, {Kozhurina-Platais}, {Larsen}, {Hollyhead}, {Lardo},
  {Cabrera-Ziri}, {Kacharov}, {Platais}, {Salaris}, {Cordero}, {Dalessandro},
  {Geisler}, {Hilker}, {Li}, {Mackey}, \& {Mucciarelli}}]{Florian:17b}
{Niederhofer} F. {et~al.}, 2017{\natexlab{a}}, MNRAS, 465, 4159

\bibitem[{{Niederhofer} {et~al}\mbox{.}(2017{\natexlab{b}}){Niederhofer},
  {Bastian}, {Kozhurina-Platais}, {Larsen}, {Salaris}, {Dalessandro},
  {Mucciarelli}, {Cabrera-Ziri}, {Cordero}, {Geisler}, {Hilker}, {Hollyhead},
  {Kacharov}, {Lardo}, {Li}, {Mackey}, \& {Platais}}]{florian121}
{Niederhofer} F. {et~al.}, 2017{\natexlab{b}}, MNRAS, 464, 94

\bibitem[{{Niederhofer} {et~al}\mbox{.}(2015){Niederhofer}, {Hilker},
  {Bastian}, \& {Silva-Villa}}]{florianAge}
{Niederhofer} F., {Hilker} M., {Bastian} N., {Silva-Villa} E., 2015, A\&A, 575,
  A62

\bibitem[{{Piotto} {et~al}\mbox{.}(2015){Piotto}, {Milone}, {Bedin},
  {Anderson}, {King}, {Marino}, {Nardiello}, {Aparicio}, {Barbuy}, {Bellini},
  {Brown}, {Cassisi}, {Cool}, {Cunial}, {Dalessandro}, {D'Antona}, {Ferraro},
  {Hidalgo}, {Lanzoni}, {Monelli}, {Ortolani}, {Renzini}, {Salaris},
  {SarAJedini}, {van der Marel}, {Vesperini}, \& {Zoccali}}]{piotto15}
{Piotto} G. {et~al.}, 2015, AJ, 149, 91

\bibitem[{{Portegies Zwart}, {McMillan} \& {Gieles}(2010){Portegies Zwart},
  {McMillan}, \& {Gieles}}]{portegies10}
{Portegies Zwart} S.~F., {McMillan} S.~L.~W., {Gieles} M., 2010, A\&ARv, 48,
  431

\bibitem[{{Reddy} {et~al}\mbox{.}(2003){Reddy}, {Tomkin}, {Lambert}, \&
  {Allende Prieto}}]{reddy03}
{Reddy} B.~E., {Tomkin} J., {Lambert} D.~L., {Allende Prieto} C., 2003, MNRAS,
  340, 304

\bibitem[{{Sbordone} {et~al}\mbox{.}(2011){Sbordone}, {Salaris}, {Weiss}, \&
  {Cassisi}}]{sbordone11}
{Sbordone} L., {Salaris} M., {Weiss} A., {Cassisi} S., 2011, A\&A, 534, A9

\bibitem[{{Schiavon} {et~al}\mbox{.}(2013){Schiavon}, {Caldwell}, {Conroy},
  {Graves}, {Strader}, {MacArthur}, {Courteau}, \& {Harding}}]{schiavon13}
{Schiavon} R.~P., {Caldwell} N., {Conroy} C., {Graves} G.~J., {Strader} J.,
  {MacArthur} L.~A., {Courteau} S., {Harding} P., 2013, ApJL, 776, L7

\bibitem[{{Schiavon} {et~al}\mbox{.}(2017{\natexlab{a}}){Schiavon}, {Johnson},
  {Frinchaboy}, {Zasowski}, {M{\'e}sz{\'a}ros}, {Garc{\'{\i}}a-Hern{\'a}ndez},
  {Cohen}, {Tang}, {Villanova}, {Geisler}, {Beers}, {Fern{\'a}ndez-Trincado},
  {Garc{\'{\i}}a P{\'e}rez}, {Lucatello}, {Majewski}, {Martell}, {O'Connell},
  {Prieto}, {Bizyaev}, {Carrera}, {Lane}, {Malanushenko}, {Malanushenko},
  {Mu{\~n}oz}, {Nitschelm}, {Oravetz}, {Pan}, {Roman-Lopes}, {Schultheis}, \&
  {Simmons}}]{schiavon17}
{Schiavon} R.~P. {et~al.}, 2017{\natexlab{a}}, MNRAS, 466, 1010

\bibitem[{{Schiavon} {et~al}\mbox{.}(2017{\natexlab{b}}){Schiavon}, {Zamora},
  {Carrera}, {Lucatello}, {Robin}, {Ness}, {Martell}, {Smith},
  {Garc{\'{\i}}a-Hern{\'a}ndez}, {Manchado}, {Sch{\"o}nrich}, {Bastian},
  {Chiappini}, {Shetrone}, {Mackereth}, {Williams}, {M{\'e}sz{\'a}ros},
  {Allende Prieto}, {Anders}, {Bizyaev}, {Beers}, {Chojnowski}, {Cunha},
  {Epstein}, {Frinchaboy}, {Garc{\'{\i}}a P{\'e}rez}, {Hearty}, {Holtzman},
  {Johnson}, {Kinemuchi}, {Majewski}, {Muna}, {Nidever}, {Nguyen}, {O'Connell},
  {Oravetz}, {Pan}, {Pinsonneault}, {Schneider}, {Schultheis}, {Simmons},
  {Skrutskie}, {Sobeck}, {Wilson}, \& {Zasowski}}]{schiavon16}
{Schiavon} R.~P. {et~al.}, 2017{\natexlab{b}}, MNRAS, 465, 501

\bibitem[{{Whitmore} {et~al}\mbox{.}(2010){Whitmore}, {Chandar}, {Schweizer},
  {Rothberg}, {Leitherer}, {Rieke}, {Rieke}, {Blair}, {Mengel}, \&
  {Alonso-Herrero}}]{whitmore10}
{Whitmore} B.~C. {et~al.}, 2010, AJ, 140, 75

\end{thebibliography}

\label{lastpage}
\end{document}